\documentclass[aps,pre,twocolumn,superscriptaddress,showpacs]{revtex4-1}

\usepackage[usenames]{color}
\usepackage{amsmath}
\usepackage{graphicx}
\usepackage{subfigure}
\usepackage{enumitem}
\begin{document}

\title{Dynamics of link states in complex networks: the case of a majority rule}

\author{J. Fern\'andez-Gracia}
\email[]{juanf@ifisc.uib-csic.es}
\affiliation{ IFISC, Instituto de F\'isica Interdisciplinar y Sistemas Complejos (CSIC-UIB), Campus Universitat Illes Balears, E-07122 Palma de Mallorca, Spain}
\author{X. Castell\'o}
\altaffiliation[Current address:]{ Departament de Psicologia Social. Universitat Aut\`onoma de Barcelona, Campus Bellaterra. 08193 Barcelona, Spain}
\affiliation{ IFISC, Instituto de F\'isica Interdisciplinar y Sistemas Complejos (CSIC-UIB), Campus Universitat Illes Balears, E-07122 Palma de Mallorca, Spain}
\author{V.M. Egu\'iluz}
\author{M. San Miguel}
\affiliation{ IFISC, Instituto de F\'isica Interdisciplinar y Sistemas Complejos (CSIC-UIB), Campus Universitat Illes Balears, E-07122 Palma de Mallorca, Spain}

\date{\today}

\begin{abstract}
% \textcolor{green}{
% \begin{itemize}
%  \item Link-based majority rule.
%  \item Node characteristics arise naturally.
%  \item Characterization of time evolution.
%  \item Characterization of heterogeneous asymptotic states.
%  \item Different networks.
%  \item Understanding in terms of the line-graph.
% \end{itemize}
% }

Motivated by the idea that some characteristics are specific to the relations between individuals and not of the individuals themselves, we study a prototype model for the dynamics of the states of the links in a fixed network of interacting units. Each link in the network can be in one of two equivalent states. A majority link-dynamics rule is implemented, so that in each dynamical step the state of a randomly chosen link is updated to the state of the majority of neighboring links. Nodes can be characterized by a link heterogeneity index, giving a measure of the likelihood of a node to have a link in one of the two states. We consider this link-dynamics model on fully connected networks, square lattices and Erd\"os-Renyi random networks. In each case we find and characterize a number of nontrivial asymptotic configurations, as well as some of the mechanisms leading to them and the time evolution of the link heterogeneity index distribution. For a fully connected network and random networks there is a broad distribution of possible asymptotic configurations. Most asymptotic configurations that result from link-dynamics have no counterpart under traditional node dynamics in the same topologies. \\

\end{abstract}

% insert suggested PACS numbers in braces on next line
\pacs{89.75.Fb,87.23.Ge,89.65.-s,89.75.Hc}
% 89.75.Fb Structures and organization in complex systems
% 87.23.Ge Dynamics of social systems
% 89.65.-s Social and economic systems
% 89.75.Hc Networks and genealogical trees
\maketitle

\section{INTRODUCTION}
Collective properties of interacting units have been traditionally studied considering that each of these units has a property or state, and interacts with others in a network of interactions. The result of the interaction depends on the state of the interacting units. For example, in a spin system in a lattice, there is a spin in each node with a given state and it interacts with its neighbors in the lattice, in a way that depends on their relative spin state. The same basic set up has been implemented in individual or agent based models of social collective properties  \cite{castellano_stat_phys}. These models endow individuals with a variable, which can be discrete or continuous, describing for example, an opinion state. The model also prescribes a dynamical rule, which results in changes of the states of the agents that depend on the state of the agents with whom they interact. However, there is a number of possible interactions among individuals in which the state variable is more properly described as a state of the interaction link than a state of the individuals in interaction. This is specially the case for relational interactions such as  friendship, trust, communication channel (phone or skype), method of salutation (kiss or handshake), etc. It is also the case in language competition dynamics \cite{Baronchelli2012}: So far language has been modeled in this context \cite{castello2006,patriarca2012} as an individual property, but use of a language, as opposed to knowledge of a language, is more a link than a node property in a social network of interactions. Noteworthy, data on link states associated with trust, friendship or enmity, obtained from on-line games and on-line communities, is now available and has been recently analyzed  \cite{Szell2010,Leskovec2010}. \\
% {\color{red}: Leskovecbis2010 es arXiv:1003.2424. No se si esta es la referencia que tu tenias como Leskovec2010} No es la misma referencia, perosi es el mismo trbajo presentado en distintas conferencias.

Social balance theory \cite{Heider1946} is a well established precedent in the study on link states and link interactions. This theory states that a relation can be positive or negative and that there is a natural tendency to avoid unbalanced triads of individuals, where an unbalanced triad is defined as one for which the multiplication of the states of the three links is negative. A number of recent studies address social balance in complex networks, implementing stochastic link dynamics that explore when a balanced situation is or is not reached asymptotically \cite{Antal2005,ANTAL2006,Radicchi2007}. Social balance theory has also been confronted with large scale data \cite{Szell2010,Leskovec2010}, and alternative theories for the interaction of positive/negative relations have also been proposed \cite{Leskovec2010,Marvel2010}. Focusing on link properties has also been emphasized in the problem of community detection in complex networks \cite{Ahn2010,Evans2009,Evans2010,Liu2010,Traag2009}. This opposes the traditional view of identifying network communities with a set of nodes \cite{Santo201075}, and it makes  possible for an individual to be assigned to different communities. Finally, the idea of considering link dynamics is also present in the problem of network dynamics controllability \cite{Nepusz2012}. Here the aim is to identify the most relevant links to drive the system to a desired global state of the network, instead of focusing on the dynamically most influent nodes \cite{Klemm2012}.\\

The aim of this paper is to investigate a prototype model for the dynamics of link states in a fixed network. Links can be in two equivalent states. This departs from the positive/negative interactions, considered for example in social balance, where the two link states play different roles. Equivalent link states can occur in many relational interactions including, for example, salutation or competition of languages of the same prestige. As a first step towards the characterization of such link dynamics we choose to investigate a majority rule dynamics akin to a zero-temperature kinetic Ising model but for the states of the links, instead of the state of the nodes. We show that such link majority rule dynamics on complex networks results in a variety of non-trivial different asymptotic configurations which are generally not found when studying traditional node-dynamics in the same topologies. We also show how a quantity characterizing the node behavior naturally arises for the link states, so that nodes can also be characterized by the state of the links connected to the node.\\

The paper is organized as follows: in section II we define our majority rule link dynamics model, as well as some quantities introduced for its characterization. In sections III, IV and V we describe our results on a fully connected network, a square lattice and Erd\"os-Renyi random networks, respectively. Section VI contains a discussion summary. Besides the material presented in this paper, which is self-explanatory, the supplementary material available in the web will help visualize different aspects of the model on different networks. \\

\section{A MODEL FOR LINK DYNAMICS}
We consider a fixed network composed by $N$ nodes and $L$ edges. The state of each link $i-j$ is characterized by a binary variable $s_{i-j}$ which can take two equivalent values $A$ or $B$. We study a majority rule for the dynamics of the state of the links. The basic step of the dynamics is
\begin{enumerate}
 \item[i.] Randomly choose a link $i-j$.
 \item[ii.] Update its state to the one of the majority of links in its first neighborhood (two links are considered first neighbors if they are attached to a common node). In case of a tie, the state of the link is randomly chosen
\end{enumerate}
The time unit is set to $N$ basic steps so that for each node, on the average, the state of one of its links is updated per unit time. This dynamics corresponds to the usual zero temperature Glauber dynamics.

 There exist two trivial absorbing ordered configurations, for which all the links in the system have the same state. The dynamics tend to order the system locally. We investigate whether, depending on the topology of the network, the dynamics orders the system globally or if the system reaches asymptotic disordered configurations with coexistence of both states. We will also characterize the transient dynamics towards these asymptotic configurations. For these purposes we will consider the following quantities characterizing the network and its links dynamics\\

\begin{description}
\item [$k_i$]{Degree of node $i$.}
 \item [$k^{A/B}_i$]{ Number of $A/B$ edges connecting node $i$. Obviously the sum of both types of links is the degree of the node, $k^{A}_i+k^{B}_i=k_i$.}
  \item [$\rho$] { Density of nodal interfaces: It serves as order parameter. It gives a local measure of the level of order in the system. Take a node of degree $k_i$. We can map it to a set of $k_i$ nodes which represent the links attached to that node. Those links are all neighbors of each other through node $i$, therefore they form a fully connected subgraph of size $k_i$. The order parameter is just the density of ties in that graph which connect links of the original network holding different states. Therefore the order parameter is zero only for the fully ordered configurations. In section \ref{discusion} we will come back to this mapping of links into nodes. $$\rho=\frac{\sum_{i=1}^Nk_i^Ak_i^B}{\sum_{i=1}^Nk_i(k_i-1)/2},$$}
 \item [$b_i$]{Link heterogeneity index of node $i$. It is a node characterization giving a measure of the likelihood of a node to have $A$ or $B$ links ($b_i=+1$ or $b_i=-1$ for all links of the same type, $b_i=0$ for a completely symmetric case). $$b_i=\frac{k_i^A-k_i^B}{k_i}.$$}
 \item [$P(b,t)$]{Link heterogeneity index distribution, probability that a randomly chosen node has link heterogeneity index $b$ at time $t$.}
 \item [$S(t)$]{Survival probability, probability that a realization of the stochastic link dynamics has not reached a fully ordered configuration at time $t$.}
\end{description}
% In the remainder we investigate, with the help of the above defined quantities, the dynamical and asymptotic properties of the model on fully connected networks, square lattices and Erd\"os-Renyi random networks.
\section{FULLY CONNECTED NETWORK}
We first consider a fully connected network in which every node is connected to every other node so that $L=N(N-1)/2$. Note however that every link is not a first neighbor of every other link.
\subsection{Time evolution}
\begin{figure}
 \centering
 \includegraphics[height=0.5\textwidth,angle=-90]{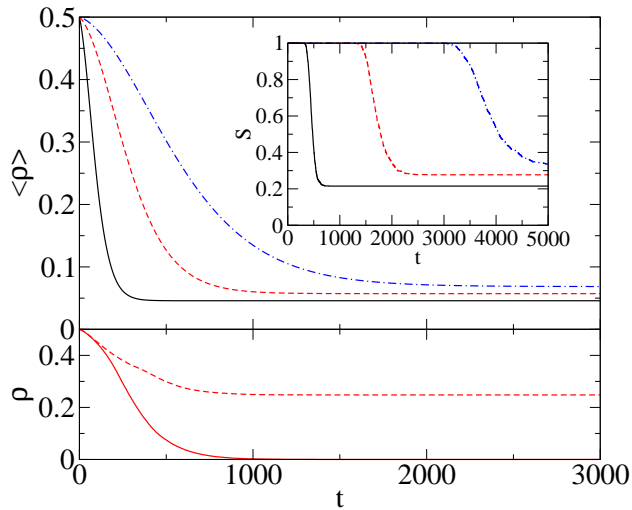}
 \caption{(Color online) Upper panel: Evolution of the average order parameter on a fully connected network. Inset: Survival probability.  $N=100$ for the black solid line, $N=300$ for the red dashed line and $N=600$ for the blue dashed-dotted line. Averages are taken over $10^3$ realizations. Lower panel: Evolution of the order parameter for single realizations of the dynamics on a fully connected network of size $N=300$. We show two different kinds of realizations: a realization reaching an absorbing ordered state (solid line) and a realization ending in a disordered frozen configuration (dashed line).\label{average_FC}}
\end{figure}

Fig.~\ref{average_FC} shows the time evolution of the ensemble average of the order parameter $\langle\rho\rangle$ and the survival probability $S(t)$ for random initial conditions. The average order parameter shows a decay towards a plateau, indicating that the absorbing ordered configurations are not always reached. Comparing this result with the survival probability, which also saturates at a certain value after a transient, we conclude that the plateau in the average order parameter is due to realizations which get frozen in a configuration with coexistence of states. The analysis of single realizations of the link dynamics (lower panel of Fig.~\ref{average_FC}) shows smooth dynamics to an asymptotic state in which the order parameter is frozen. In the following we investigate the characteristics of these frozen asymptotic configurations. \\

\subsection{Asymptotic configurations}
\begin{figure}
 \centering
 \includegraphics[height=0.5\textwidth,angle=-90]{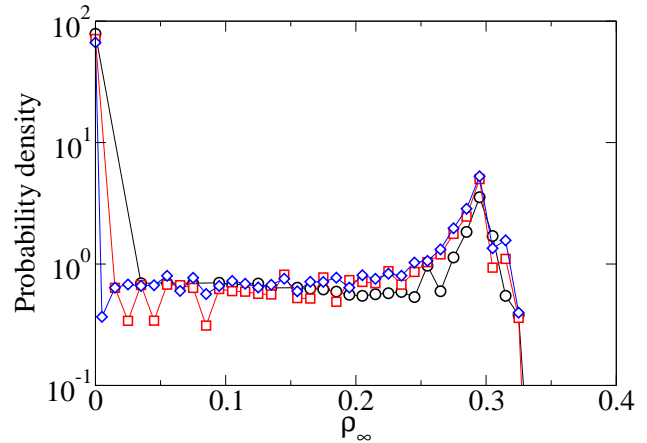}
 \caption{(Color online) Probability density of the asymptotic value of the order parameter $\rho_{infty}$ for a complete graph. The calculation is done over $10^4$ realizations for system sizes $N=100$ (black circles), $N=300$ (red squares) and $N=600$ (blue diamonds).\label{CG_final_rho}}
\end{figure}

We checked that all asymptotic configurations are frozen, \emph{i.e.} the order parameter $\rho$ and the densities of links in each state reach a value that stays constant over time. The probability of having a certain value of $\rho$ in the asymptotic configurations, $\rho_{\infty}$, is plotted in Fig.~\ref{CG_final_rho}. We observe a very heterogeneous set of possible final configurations in addition of the most probable ordered configuration ($\rho=0$). For classifying the disordered frozen configurations we use the number $n_b$ of elements in the set of link heterogeneity indices present in each configuration. The limiting case $n_b=1$ corresponds to the ordered configuration in which all nodes have the same heterogeneity index $b=1$ or $b=-1$. The case $n_b=2$ corresponds to a family of asymptotic configurations where the number of different link heterogeneity indices of the nodes is two, and therefore the nodes can be divided into two groups. These configurations are depicted in Fig.~\ref{cg_frozen}.a.
\subsubsection{Simplest frozen configurations}
\begin{figure}
 \centering
 \includegraphics[scale=0.6]{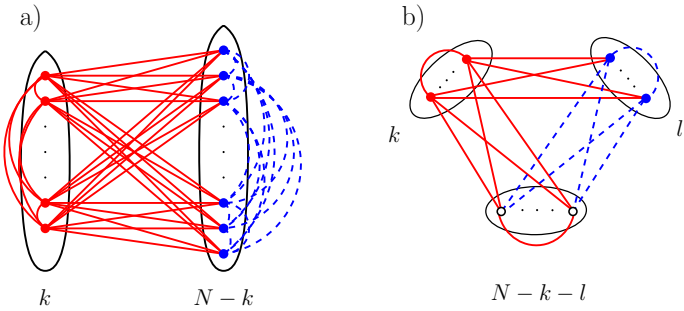}
 \caption{(Color online) a) Simple frozen configurations in a fully connected network. b) Frozen configuration with $n_b=3$ on a fully connected network. The states of the links are encoded by color and line style, so there are red solid lines and blue dashed lines representing both states.\label{cg_frozen}}
\end{figure}

The simplest frozen configurations on a fully connected network are of the type shown in Fig.~\ref{cg_frozen}a. They consist of a set of $k$ nodes that have only links in one state (red links), and the rest of nodes, $N-k$, having all their links in the other state (blue links), except for the links with the $k$ nodes of the first set. It is clear that for this kind of configuration there are only two types of nodes in terms of link heterogeneity index. The group of size $k$ having $|b|=1$ and therefore contributing to the asymptotic $p(b,t=\infty)$ in the peaks $b=\pm1$ (see Fig.\ref{evol_average_pdeb} a), and another group of size $N-k$ with $|b|=(2k-N-1)/(N-1)$. Therefore, for these configurations $n_b=2$.\\

\begin{figure}
 \centering
 \includegraphics[height=0.5\textwidth,angle=-90]{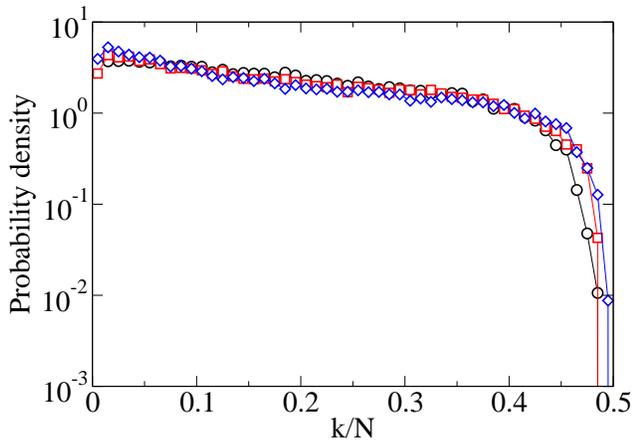}
 \caption{(Color online) Probability density of getting to a simple frozen configuration like the one in Fig.~\ref{cg_frozen}.a) with a certain fraction $k/N$ of nodes with $|b|=1$, starting from random initial conditions on a complete graph. Sizes are $N=100$ (black circles), $N=300$ (red squares) and $N=600$ (blue diamonds). The statistics are over $10^5$ realizations of the system.\label{monofraction}}
\end{figure}

These configurations can be proven to be frozen for a range of sizes $k$. For this purpose one has to find how many of the neighboring links of a given link are in each state and impose that links in state A (B) have more A (B) neighbors than B (A) neighbors. In this way one easily concludes that configurations as the one in Fig.~\ref{cg_frozen}.a) are frozen whenever $$1<k<N/2-1.$$ These solutions exist for $N>4$. In Fig.~\ref{monofraction} we show the probability density to reach a configuration with a certain fraction $k/N$ with $|b|=1$, given that the asymptotic configuration is of the type shown in Fig.~\ref{cg_frozen}a. All the possible configurations can be reached from random initial conditions.

\subsubsection{Other asymptotic configurations}
\begin{figure}
 \centering
 \includegraphics[width=0.5\textwidth]{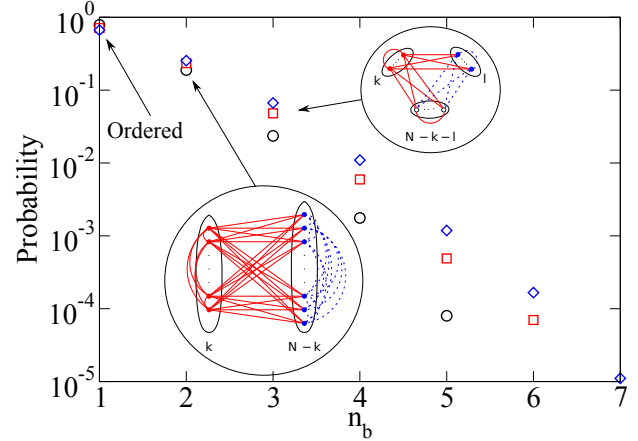}
 \caption{(Color online) Probability of reaching a frozen configuration with a certain number of different link heterogeneity indices $n_b$, starting from random initial conditions on a complete graph. Sizes are $N=100$ (black circles), $N=300$ (red squares) and $N=600$ (blue diamonds), and the statistics are over $10^5$ realizations of the system.\label{numbil}}
\end{figure}

 Figure~\ref{numbil} shows the probability of reaching an asymptotic configuration with a certain number of different link heterogeneity indices $n_b$ in the system. The ordered configurations $n_b=1$ and the ones with $n_b=2$ described above are the most probable. An example of a configuration with $n_b=3$ is shown in Fig.~\ref{cg_frozen}b. These configurations have $k$ nodes with $|b|=1$, $l$ nodes with $b=(2k-N+1)/(N-1)$ and $N-k-l$ nodes with $b=(N-2l-1)/(N-1)$. Following the same argument used for $n_b=2$ configurations, we can conclude that $n_b=3$ configurations are frozen provided that\\
 \begin{align}
 k&>1\nonumber \\
 l&<N/2-1\nonumber \\
 k&<N/2-1\nonumber \\
 l&>k+1\nonumber
\end{align}
When $n_b$ is increased, the frozen configurations become structurally more complicated, and are much less probable. Empirically we have found that always a group of agents with $|b|=1$ appears. To characterize a frozen solution with $n_b$ we need $n_b-1$ parameters and $n_b(n_b+1)/2$ inequalities, which arise  imposing that the state of every link is frozen and give a boundary for the possible architecture of those configurations.\\

% \begin{figure}
%  \centering
%  \includegraphics[scale=0.6]{frozen_CG_nb3}
%  \caption{Frozen configuration with $n_b=3$ in a fully connected network. We have three kinds of nodes that we will call red nodes, blue nodes and white nodes.\label{cg_frozen_nb3}}
% \end{figure}
% \emph{Conditions for being frozen.}

\subsection{Link heterogeneity index distribution}
\begin{figure}
 \centering
 \includegraphics[height=0.5\textwidth,angle=-90]{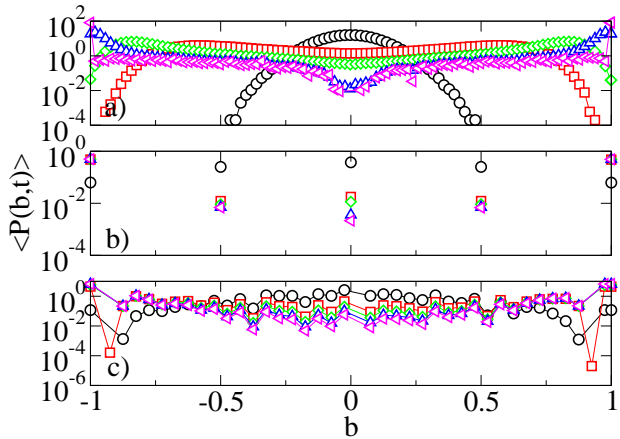}
 \caption{(Color online) Probability density distribution of link heterogeneity index $P(b,t)$ for different times averaged over $10^3$ realizations. The initial condition is in black circles. Time evolution is in the following order:  red squares, green diamonds, blue up triangles and magenta left triangles. a) Fully connected network of size $N=100$ at times $0,50,100,200,500$. b) Square lattice of size $N=2500$ for times $0,500,1000,2000,3000$. In this case the actual probability of having index $b$ is plotted and not the density. c) Erd\"os-Renyi random network of size $N=1000$ and average degree $\langle k\rangle=10$ at times $0,50,100,200,500$. All plots are symmetric around $b=0$ due to the equivalent nature of the states A and B. Note the logarithmic scale in the y-axis and therefore that the peaks at $\pm1$ are much higher than the rest of the distribution for long times.\label{evol_average_pdeb}}
\end{figure}

% {\color{red}:Ganamos algo poniendo en esta figura las disitntas redes juntas o mejor separalo en varias figuras como en la presentacion ppt?}
Figure~\ref{evol_average_pdeb}.a shows the time evolution of the link heterogeneity index distribution: We observe that it evolves from a distribution peaked around $b=0$ for random initial conditions, to a bimodal distribution peaked around $b=\pm1$, with a quite homogeneous probability of having any link heterogeneity index. This statistics characterized by this distribution includes the most probable realizations that reach ordered states but also others which freeze in configurations with nodes with different possible values of $b$, as discussed in the characterization of the asymptotic configurations. Note that both type of realizations contribute to the peaks at $b=\pm1$  since frozen disordered asymptotic configuration have at least one group of agent with $b=\pm1$.\\
% \begin{figure}
%  \centering
%  \includegraphics[height=0.5\textwidth,angle=-90]{evol_single_pdeb}
%  \caption{Same as in Fig.~\ref{evol_average_pdeb} for single realizations of the system.\label{evol_single_pdeb}}
% \end{figure}
\section{SQUARE LATTICE}
In order to account for local interaction effects we first consider a square lattice with periodic boundary conditions. We used only square lattices with a square number of sites, \emph{i.e.} $N=l^2$ with $l$ being the length of the side of the network.
\subsection{Time evolution}
\begin{figure}
 \centering
 \includegraphics[height=0.5\textwidth,angle=-90]{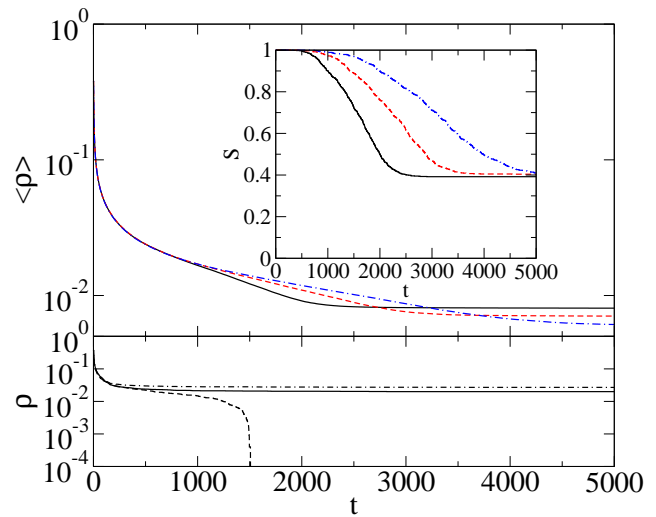}
 \caption{(Color online) Upper panel: Evolution of the average order parameter on a square lattice. Inset: Survival probability.  $N=2500$ for the black solid line, $N=3600$ for the red dashed line and $N=4900$ for the blue dashed-dotted line.  Averages are taken over $10^3$ realizations. Lower panel: Evolution of the order parameter for single realizations of the dynamics on a square lattice of size $N=2500$. We show three different realizations, corresponding to the three possible asymptotic configurations: ordered state (dashed line), vertical/horizontal single stripe (solid line) and diagonal single stripe (dotted-dashed line).\label{average_2D}}
\end{figure}

Fig.~\ref{average_2D} shows the time evolution the ensemble average of the order parameter $\langle\rho\rangle$ and the survival probability $S(t)$ for random initial conditions. We observe a qualitative behavior very similar to the one in Fig.~\ref{average_FC} for a fully connected network, \emph{i.e.}  $\langle \rho\rangle$ and $S(t)$ decay smoothly to a plateau value. Together with the plot of single realizations of the stochastic dynamics (lower panel of Fig.~\ref{average_2D}~) this indicates that some of the realizations reach an asymptotic ordered state, while others get trapped in a disordered configuration for which the order parameter remains constant for all times. We have found only three different types of realizations characterized by their asymptotic configurations, as we discuss next.\\
\subsection{Asymptotic configurations}
\begin{figure}
 \centering
 \includegraphics[width=0.5\textwidth]{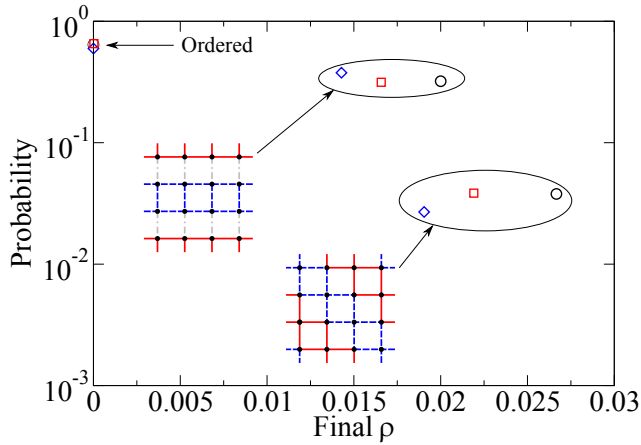}
 \caption{(Color online) Probability of reaching a given asymptotic value of the order parameter on a square lattice with periodic boundary conditions and starting from random initial conditions. There are three different possible configurations, namely ordered state, horizontal/vertical stripes and diagonal stripes. Sizes are $N=2500$ (black circles), $N=3600$ (red squares) and $N=4900$ (blue diamonds). Statistics computed from $10^4$ realizations. \label{square_final_rho}}
\end{figure}

\begin{figure}
 \centering
 \includegraphics[width=0.5\textwidth]{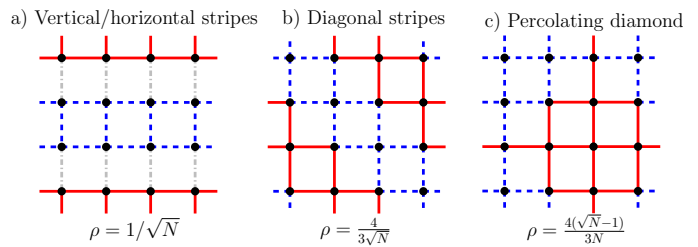}
 \caption{(Color online) Different asymptotic disordered configurations on a square lattice with periodic boundary conditions. a) Vertical/horizontal single stripe. The gray dash-dotted links keep changing state forever, while all other links are in a frozen state. b) Diagonal single stripe. All links are frozen. c) Percolating diamond. All links are frozen.\label{square_asymp}}
\end{figure}

The probability of reaching one of the three main possible asymptotic configurations, characterized by their value of the order parameter, is shown in Fig.~\ref{square_final_rho}. These configurations are depicted in Fig.~\ref{square_asymp}. Note that these are the asymptotic configurations starting from random initial conditions and using periodic boundary conditions.

\begin{itemize}
 \item \emph{Ordered configurations:} All links are in the same state and $\rho=0$.
 \item \emph{Dynamically trapped configurations}, where the order parameter remains constant,  $\rho=1/\sqrt{N}$, but the densities of links in each state fluctuate around a certain value. These configurations form vertical/horizontal stripes, as shown in Fig.~\ref{square_asymp}a. These configurations are dynamical traps from which the system cannot reach the ordered state: links in the boundaries of the stripes continue to blink without changing the value of the order parameter. Single stripe are the configurations reached from random initial conditions. However configurations with a larger number of stripes ( and thus a value of the order parameter which is a multiple of $1/\sqrt{N}$) are also dynamical traps of the model.
 \item \emph{Frozen configurations}, where the order parameter and the densities of links in each state remain constant. Configurations reached from random initial conditions are single diagonal stripe as the one shown in Fig.~\ref{square_asymp}b, with a value of the order parameter $\rho=\frac{4}{3\sqrt{N}}$. There are other frozen configurations for our dynamical model which we have not observed in our simulations with random initial conditions. These include multiple diagonal stripes and a combination of diagonal front that we call percolating diamond (Fig.~\ref{square_asymp}c): It contains a square of links in one state, rotated an angle of 45 degrees, surrounded by links in the opposite state and which percolates through the network. For the percolating diamond configuration  $\rho=\frac{4(\sqrt{N}-1)}{3N}$.
\end{itemize}

Note that given an asymptotic configuration and the size of the network the value of the order parameter $\rho$ can be known and is the same independently of the width of the stripes in the stripe patterns, as the order parameter only contributes in the boundaries of the stripes.

\subsection{Link heterogeneity index distribution}
For a square lattice the link heterogeneity index takes values $b=\pm1, \pm0.5, 0$. The evolution of the distribution $P(b,t)$  is shown in Fig.~\ref{evol_average_pdeb}b.  It evolves from an initial peak at $b=0$ to a distribution with two peaks at $b=\pm1$ and a minimum value at $b=0$. This evolution can be understood from the asymptotic configurations described above: The two peaks at $b=\pm1$ originate in the most probable ordered configurations, but also on the large percentage of nodes with $b=\pm1$ in the two other possible asymptotic configurations. The values $b=\pm0.5$ appear only in the second most probable asymptotic configuration: vertical/horizontal single stripe. For these configurations there are $4\sqrt{N}$ nodes whose heterogeneity index keeps jumping from $b=\pm1$ and $b=\pm0.5$. Last, the probability of having nodes with $b=0$ comes from the third possible asymptotic configuration, diagonal single stripe. In this configuration $2\sqrt{N}-2$ nodes have $b=0$ and the other nodes are divided into two equal groups with $b=1$ and $b=-1$.

\section{RANDOM NETWORKS}
In order to account for the role of network heterogeneity we finally consider the link dynamics model on  Erd\"os-Renyi random networks.
\subsection{Time evolution}
\begin{figure}
 \centering
 \includegraphics[height=0.5\textwidth,angle=-90]{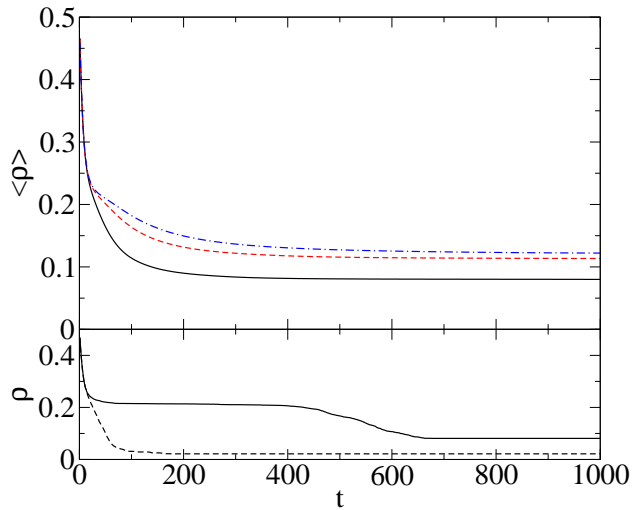}
 \caption{(Color online) Upper panel: Evolution of the average order parameter on Erd\"os-Renyi networks of average degree $\langle k\rangle=10$.  $N=1000$ for the black solid line, $N=5000$ for the red dashed line and $N=10000$ for the blue dashed-dotted line. Averages are taken over $10^3$ realizations of different initial conditions and different realizations of the random network. Lower panel: Evolution of the order parameter for single realizations of stochastic dynamics on an Erd\"os-Renyi random network of size $N=1000$ and average degree $\langle k\rangle=10$. Two different realizations are shown, each one ending in a different configuration with frozen order parameter.\label{average_ERm5}}
\end{figure}

Proceeding as in the previous cases we show in Fig.~\ref{average_ERm5} the time evolution of the ensemble average order parameter. The survival probability (not shown) is one at all times,  except for small systems or networks of high average degree where it tends to a fully connected like behavior. Our results indicate that all stochastic realizations of the dynamics reach an asymptotic disordered configuration with a constant value of $\rho$.
\subsection{Asymptotic configurations}
\begin{figure}
 \centering
 \includegraphics[height=0.5\textwidth,angle=-90]{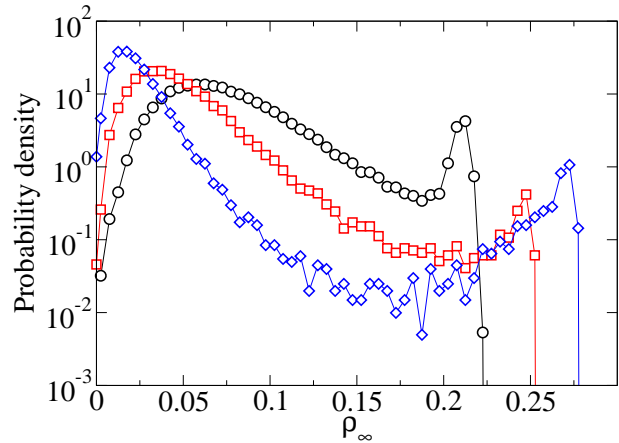}
 \caption{(Color online) Probability density of the asymptotic value of the order parameter $\rho_{infty}$ for a random graph. The calculation is done over $10^4$ realizations for system size $N=1000$ and average degrees $\langle k\rangle=10$ (black circles), $\langle k\rangle=20$ (red squares) and $\langle k\rangle=40$ (blue diamonds).\label{ER_final_rho}}
\end{figure}

  We observe a large variety of asymptotic configurations characterized by different values of the order parameter $\rho$, as shown in Fig.~\ref{ER_final_rho}. Increasing the average degree of the network, the distribution of final values of  $\rho$ approaches the one for a fully connected network (Fig.~\ref{CG_final_rho}): The distribution develops a peak that moves towards $\rho=0$ and another peak near the maximum asymptotic order parameter value that is accessible starting from random initial conditions.\\

For random networks we also find three kinds of asymptotic configurations:
\begin{itemize}
 \item \emph{Ordered configurations:} All links are in the same state and $\rho=0$. This configuration is only observed in small systems or in systems with high average degree (close to fully connected network).
 \item \emph{Dynamically trapped configurations:} the order parameter remains constant, but the densities of links in each state vary with time.
 \item \emph{Frozen configurations}, where the order parameter and the densities of links in each state remain constant.
\end{itemize}

\begin{figure}
 \centering
 \includegraphics[scale=0.6]{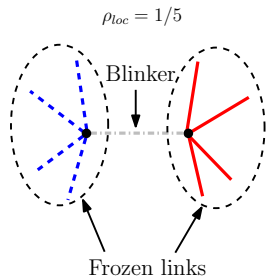}
 \caption{(Color online) Example of change of state which changes the densities of blue and red links but conserves the value of the order parameter $\rho$. Independently of the state of the grey link this motif will contribute to the order parameter of the whole system with $\rho=1/5$.\label{isoenergetic2}}
\end{figure}

\begin{figure}
 \centering
 \includegraphics[width=0.49\textwidth]{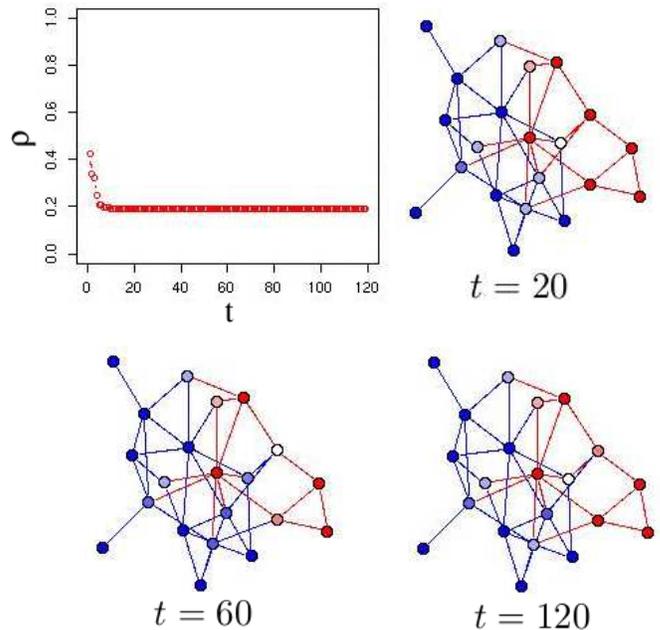}
 \caption{(Color online) One realization on a small random network of size $N=20$. Top left panel shows the evolution of the order parameter, which freezes after approximately $10$ time steps. The other panels show the configuration of the system at different times. The color of the nodes reflects their link heterogeneity index. Red (blue) is for having all links in the red (blue) option, white is for having half of the links in each color. The changes in the configuration do not affect the value of the order parameter. For example the only difference between the configuration at $t=20$ and the one at $t=120$ is the state of a single link. If we count we can see that the link has the same number of neighbors in each state. One can check that all the changes of state are of the type depicted in Fig.~\ref{isoenergetic2}\label{animation}}
\end{figure}

It is possible to identify some basic mechanisms leading to the observed traps. Among them:
\begin{itemize}
\item \emph{The role of hubs:}
If a node $i$ is such that $k_i\gg k_j$ for any neighboring node $j$, then the links attached to node $i$ usually end up sharing all the same state, which in most cases is the one of the initial majority state in that set of links. This effect creates \emph{frozen} links, \emph{i.e.} links which do not change state. Initial conditions and the particular topology of the realization will determine how frequent is this effect and whether this leads or not to an ordered configuration. 
\item \emph{Dynamics conserving the value of $\rho$:}
There exist changes of the state of the links which do not cause a change in the value of $\rho$. These changes are those for which the link changing  state has a symmetric environment, with the same number of neighbors in each state as shown in Fig.~\ref{isoenergetic2}. This situation is the one also found in a square lattice Fig.~\ref{square_asymp}.a. This kind of phenomenon can appear in more complex forms, as shown in Fig.~\ref{animation}. There one can see that the order parameter is frozen after approximately $10$ time steps, but the configuration of the system keeps changing, as can be seen from the snapshots of the system configuration at different times.

 %A simple way of showing that is making use of a mapping between the original network and the its line-graph. The line-graph is a network where the nodes represent the links of the original network and are connected to those nodes that represent neighboring links in the original graph. In the line-graph the order parameter is just the density of interfaces, \emph{i.e.} the number of links connecting nodes in different states. See Fig.~\ref{isoenergetic2}.\\
\end{itemize}
This behavior of the model was already pointed out in Ref.~\cite{redner_ising3d} for the Ising model in three dimensions, where the system wanders ad infinitum on a connected set of equal-energy blinker states. In Ref.~\cite{castello_epl} it was also found how community structure can induce topological traps just as is the case in the present work.


\subsection{Link heterogeneity index distribution}
The evolution of the distribution of link heterogeneity indices in random networks is shown in Fig.~\ref{evol_average_pdeb}c. The initial distribution is broad, but smoothly peaked around $b=0$. This evolves to a bimodal distribution peaked around $b=\pm1$. The hubs of the network are prone to become nodes with $b=\pm1$, which in turn pulls more nodes to this value of $b$. The fact that links can be in frozen states for different parts of the network implies that between patches of ordered \emph{domains}, there are nodes with any value of the link heterogeneity index. This contributes to the almost flat distribution of $b$ values between the two peaks. Blinking links in dynamically trapped configurations also contribute to the broad distribution of intermediate values of $b$.
\section{SUMMARY and DISCUSSION}\label{discusion}

The study of a majority rule for the dynamics of two equivalent link states in a fixed network uncovers a set of non-trivial asymptotic configurations which are generally not present when studying the classical node-based majority rule dynamics. The characterization given of the asymptotic configurations in fully connected networks, square lattices and Erd\"os-Renyi random networks provides the basis for the understanding of the evolution of the link heterogeneity index distribution. For a fully connected network and for a square lattice we have fully characterized the asymptotic configurations reached from random initial conditions. In a fully connected network we have found large heterogeneity in the asymptotic configurations. All these configurations, classified by the number $n_b$ of heterogeneity indexes present in the configuration, are frozen. Note that for the corresponding node-dynamics in the same network only an asymptotic ordered configuration is found ($n_b=1$). In a square lattice we have found asymptotic configurations which are ordered, frozen and disordered, or dynamically trapped. The latter do not have an analog in the corresponding node dynamics. In the case of Erd\"os-Renyi random networks we have described the mechanisms leading to the existence of very heterogeneous asymptotic configurations which are either frozen or dynamical traps. \\

It is clear that a link-dynamics model can be mapped into an equivalent node-based problem by changing the network of interaction. The node-equivalent network is the line-graph \cite{Krawczyk2010,Manka-Krason2009,Rooij1965} of the original network. The line-graph is a network where the links of the original network are represented by a node and are connected to those nodes that represent links that were first neighbors in the original network. This mapping of the problem has not been pursued here since it obscures our original motivation and, given the complexity of the line graphs of the networks considered here, it has been found not to be particularly useful for a quantitative description of the dynamics. However, the mapping does provide additional qualitative understanding of our findings: The line graph is a network with higher connectivity since all links that converged originally in a node form a fully connected subgraph in the line-graph, as clearly seen in the line-graph of a fully connected network or a square lattice. This results in an increased cliqueness of the line graph, as compared to the original network. Such cliqueness is behind the topological traps that give rise to the wide range of possible asymptotic configurations that we find for the link-dynamics. In addition, the mapping of a hub of the original network in the corresponding line-graph also helps understanding the different role played by hubs in node or link-based dynamics: as discussed in Section IV, hubs tend to freeze link states in their neighborhood.\\

The link heterogeneity index is a useful way of characterizing nodes in a given link configuration. For example in node based models of language competition, a node can be in state $A$ or $B$ corresponding to two competing languages, and bilingualism can only be introduced through a third node bilingual state $AB$ \cite{castello2006}. In the framework of link dynamics, state $A$ or $B$ characterizes the language used in a given interaction between two individuals, and the link heterogeneity index is a natural measure of the degree of bilingualism of each individual (node). Continuing with this example, a next step is to consider the mixed dynamics of language competence (node dynamics) and language use (link dynamics). In general, consideration of the coevolution of link and node states is a natural framework that emerges in the study of collective behavior of interacting units. In physical terms, the states of the interacting particles are coupled to the state of the field that carries the interaction. Another possible avenue of research is the addition of more realistic features to the model, such as the temporal patterns of human interactions \cite{juanfpre,tempnets}, which introduces heterogeneity in the activation of different links. 
\\

\section*{ACKNOWLEDGEMENTS}
We acknowledge financial support from the MINECO (Spain) and FEDER (EU) through projects FISICOS (FIS2007-60327) and MODASS (FIS2011-24785). J.F.-G. acknowledges a predoctoral fellowship from the Government of the Balearic Islands through the Conselleria d'Educaci\'o, Cultura i Universitats with funding from the ESF. X.C. acknowledges the Juan de la Cierva programe from the Spanish Government.

% \bibliography{LinksResponse}

%

\end{document}